\documentclass[prd,twocolumn,showpacs,preprintnumbers,floats,floatfix,nofootinbib]{revtex4-1}
\usepackage{graphicx}
\usepackage{dcolumn}
\usepackage{latexsym}
\usepackage{amsmath}
\usepackage{float}
\usepackage[caption = false]{subfig}

\def\cao{\c c\~ao}

\begin{document}

\title{Galerkin-Collocation domain decomposition method for arbitrary binary black holes}

\author{W. Barreto}
\affiliation{Centro de F\'\i sica Fundamental, Universidad de Los Andes, M\'erida 5101, Venezuela}
\affiliation{Departamento de F\'\i sica Te\'orica, Instituto de F\'\i sica A. D. Tavares,
Universidade do Estado do Rio de Janeiro, R. S\~ao Francisco Xavier, 524,
Rio de Janeiro 20550-013, RJ, Brasil}

\author{P. C. M. Clemente, H. P. de Oliveira}
\affiliation{Departamento de F\'\i sica Te\'orica, Instituto de F\'\i sica A. D. Tavares,Universidade do Estado do Rio de Janeiro, R. S\~ao Francisco Xavier, 524,
Rio de Janeiro 20550-013, RJ, Brasil}

\author{B. Rodriguez-Mueller}
\affiliation{Computational Science Research Center, San Diego State University, United States of America}

\date{\today}

\begin{abstract}
We present a new computational framework for the Galerkin-collocation method for double domain in the context of ADM 3+1 approach in numerical relativity. This work enables us to perform high resolution calculations for initial sets of two arbitrary black holes. We use the Bowen-York method for binary systems and the puncture method to solve the Hamiltonian constraint. The nonlinear numerical code solves the set of equations for the spectral modes using the standard Newton-Raphson method, LU decomposition and Gaussian quadratures. We show convergence of our code for the conformal factor and the ADM mass. Thus, we display features of the conformal factor for different masses, spins and linear momenta. 
\end{abstract}

\maketitle

\section{Introduction}%

The recent direct observation of gravitational waves by the LIGO-Virgo consortium \cite{ligo1,ligo2,ligo3,ligo4} represents an enormous breakthrough the researchers have pursued for decades. The observed wave template was generated by a binary black hole (BBH) system as predicted by the numerical simulations. In fact, Numerical General Relativity was crucial after a long effort of theoretical developments towards the obtention of a stable full dynamics of a BBH \cite{pretorius,campanelli,baker}. In this context, it is necessary a precise determination of the initial spatial hypersurface which contains the desired astrophysical configuration.

We report here a new domain decomposition algorithm (DD) based on the Galerkin-Collocation (GC) method \cite{GC_method} to obtain general initial data for a BBH system. In the present version, we are going to restrict ourselves to the Bowen-York initial data \cite{bowen-york} with the puncture \cite{brandt} wormhole foliations representing the black holes, but it can be extended to the case of puncture trumpet representation \cite{baum_naculich,hannam_09}. Albeit the existence of other spectral DD codes due to Grandclement et al. \cite{grandclement}, Pfeiffer \cite{pfeiffer,pfeiffer_CPC}, Ansorg \cite{ansorg_05,ansorg_07} and Ossokine et al.\cite{ossokine}, we believe that the present approach is a viable and valid alternative to the established codes. In the sequence, we present unique aspects of the GC-DD method that makes it structurally simple and at the same time accurate. 

The present GC-DD algorithm is a direct extension of the single domain scheme \cite{clemente} we have developed recently to describe the initial data for single and binary black holes punctures in the wormhole or trumpet representations. We highlight some of the distinct aspects of the GC domain decomposition algorithm. The basis functions are established such that each component satisfies the boundary conditions in each subdomain. We have introduced two subdomains covered by the standard spherical coordinates $(r,\theta,\phi)$ designated by $\mathcal{D}_1: 0 < r \leq r_0$ and $\mathcal{D}_2: r_0 \leq r < \infty$. In this scheme, the angular coordinates of the collocation points are common to both subdomains, although we can to chose different numbers of collocation points in each domain. We have selected the spherical harmonics as the angular basis functions. Another distinct feature of the algorithm is the particular way we have compactified the spatial domain (cf. Fig. 1).

We have organized the paper as follows. In Section II, we have briefly described the main aspects of the initial data construction of spinning-boosted binaries of black holes. Section III deals with the essential features of the GC-DD method to solve the Hamiltonian constraint. The numerical implementation of the code is described in Section IV. We have presented in Section V the validation of the algorithm with three examples of binary systems. The first and the second are equal masses boosted binary black holes in the axisymmetric and three-dimensional configurations, respectively. Whereas in the last example we have treated a more general binary of spinning-boosted black holes of different puncture masses. In all cases, we have exhibited the convergence tests.  We summarize to conclude, and we discuss of possible applications of the algorithm considered here.

\section{Basic equations}%

In General Relativity the initial data problem deals with the characterization of the gravitational and matter fields in a given initial spatial hypersurface. Equivalently, this task entails the establishment of an initial hypersurface containing plausible astrophysical systems such as binary black holes, binary of neutron stars or a binary formed by a neutron star and a black hole. According to the 3+1 formulation of the General Relativity \cite{adm,gourgoulhon}, the spacetime is foliated by a family of spatial slices $\Sigma$. Assuming the absence of matter fields, we have to solve the Hamiltonian and momentum constraint equations for $\gamma_{ij}, K_{ij}$ - the 3-metric and the extrinsic curvature associated with the initial slice, respectively - after providing their corresponding freely specifiable components.

The Hamiltonian and momentum constraints have the following forms that encompass the above requirements \cite{cook}

\begin{eqnarray}
8\bar \nabla^2 \Psi -\Psi\bar R -\frac{2}{3}\Psi^5 K^2 + \Psi^{-7}\bar A_{ij} \bar A^{ij} = 0, \label{eq1} \\
\nonumber \\
\bar{\nabla}_j \bar{A}^{ij}-\frac{2}{3}\Psi^6 \bar{\gamma}^{ij} \nabla_j K=0, \label{eq2}
\end{eqnarray}

\noindent where $\bar{\gamma}_{ij}$  is the known spatial background metric that is related to $\gamma_{ij}$ through the conformal transformation

\begin{equation}
\gamma_{ij} = \Psi^4 \bar{\gamma}_{ij}.\label{eq3}
\end{equation}  

\noindent Then, all barred quantities are related to the background metric and

\begin{equation}
A_{ij} = \Psi^{-2} \bar{A}_{ij} \label{eq4}
\end{equation}  

\noindent is the traceless part of the extrinsic curvature related to its counterpart of the background metric. 

The simplest choice for the background spatial metric is $\bar{\gamma}_{ij}=\eta_{ij}$ and together with the maximal slicing condition, $K=0$, provide the decoupling of Eqs. (\ref{eq1}) and (\ref{eq2}). As a consequence, the momentum constraint becomes a linear equation allowing to obtain the exact solutions for the components of $\bar{A}_{ij}$ corresponding to spinning and boosted black holes. This scheme characterizes the well known Bowen-York initial data \cite{bowen-york}. 

To describe binary black holes, we have considered the puncture method \cite{brandt} in which the singularities present in the conformal factor are described analytically by representing each black hole in the wormhole or trumpet slices. In the first case, the ansatz for the conformal factor is

\begin{equation}
\Psi = 1 + \frac{1}{2}\left(\frac{m_1}{r_{C_1}}+\frac{m_1}{r_{C_2}}\right) + u, \label{eq5} 
\end{equation}

\noindent where $m_1$ and $m_2$ are the puncture masses, $r_{C_i} = |\mathbf{r}-\mathbf{C}_i|$ denotes the coordinate distance to the center of the black hole located at $\mathbf{r}=\mathbf{C}_i$ and $u$ is a regular function determined after solving the Hamiltonian constraint. 

In the single domain Galerkin-Collocation algorithm \cite{clemente}, we have adopted spherical coordinates $(r,\theta,\phi)$ to cover the whole spatial domain. In the present two domain approach, we have used the same spherical coordinates in both domains instead of alternative coordinate systems as in Refs. \cite{pfeiffer,pfeiffer_CPC,ansorg_05,ansorg_07}. In this case, the regular function $u=u(r,\theta,\phi)$ satisfies the Robin boundary condition

\begin{equation}
u(r,\theta,\phi) = \mathcal{O}(r^{-1}), \label{eq6}
\end{equation}

\noindent for a large distance from the binary.

After substituting the conformal factor given by Eq. (\ref{eq5}) into the Hamiltonian constraint, we obtain

{\small
\begin{eqnarray}
\frac{1}{r^2}\frac{\partial}{\partial r}\left(r^2 \frac{\partial u}{\partial r}\right) &+& \frac{1}{r^2 \sin \theta}\frac{\partial}{\partial \theta}\left(\sin \theta \frac{\partial u}{\partial \theta}\right) + \frac{1}{r^2 \sin^2 \theta}\frac{\partial^2 u}{\partial \phi^2} \nonumber\\
\nonumber \\
 &+& \frac{\bar{A}^{ij}\bar{A}_{ij}}{8 \left[1+\frac{1}{2}\left(\frac{m_1}{r_{C_1}}+\frac{m_2}{r_{C_2}}\right) + u \right]^7} = 0. \label{eq7}
\end{eqnarray}
}

\noindent The first three terms correspond to the Laplacian of the function $u$ in spherical coordinates, and $\bar{A}^{ij}\bar{A}_{ij}$ depends upon the black holes have linear and angular momenta. Due to the linearity of the momentum constraint equation, the total background extrinsic curvature corresponding to an arbitrary binary black hole is 

\begin{eqnarray}
\bar{A}^{ij} = \bar{A}^{ij}_{\mathbf{C}_1\mathbf{P}_1} + \bar{A}^{ij}_{\mathbf{C}_1\mathbf{S}_1} + \bar{A}^{ij}_{\mathbf{C}_2\mathbf{P}_2} + \bar{A}^{ij}_{\mathbf{C}_2\mathbf{S}_2}, \label{eq8}
\end{eqnarray}

\noindent where $\bar{A}^{ij}_{\mathbf{C}_k\mathbf{P}_k}$ and $\bar{A}^{ij}_{\mathbf{C}_k\mathbf{S}_k}$ correspond, respectively, to the background extrinsic curvature of the puncture located at $\mathbf{r}=\mathbf{C}_k$, $k=1,2$, carrying linear momentum $\mathbf{P}_{k}$ and spin $\mathbf{S}_{k}$. For the sake of completeness, we have \cite{bowen-york,baumgarte}

\begin{eqnarray}
\bar{A}^{ij}_{\mathbf{C}_k\mathbf{P}_k} &=& \frac{3}{2 r_{C_k}} \left[2P^{(i}_{(k)} n^{j)}_{(k)}-(\eta^{ij}-n^i_{(k)}n^j_{(k)}) \mathbf{n}.\mathbf{P}\right] \label{eq9}\\
\nonumber \\
\bar{A}^{ij}_{\mathbf{C}_k\mathbf{S}_k} &=& \frac{6}{r_{C_k}^3} n^{(i}_{(k)}\epsilon^{j)}_{mp}S^m_{(k)}n^p_{(k)},\label{eq10}
\end{eqnarray}

\noindent where $k=1,2$ indicate each black hole and $\mathbf{n}_{k}=(\mathbf{r}-\mathbf{C}_k)/r_{C_k}$ is the normal vector to $\mathbf{r}_{C_k}$.

To complete this section we introduce the ADM mass for the arbitrary binary black holes \cite{baumgarte}

\begin{equation}
M_{ADM}=-\frac{1}{2\pi}\int_{\partial \Sigma_\infty} d\bar S_i \bar\nabla^i\psi, \label{eq11}
\end{equation}

\noindent where $\partial \Sigma_\infty$ is a surface at infinity on the spacelike foliation $\Sigma$; $d\bar S_i$ is an outward surface element. By assuming spherical coordinates and the conformal factor given by Eq. (\ref{eq5}), we obtain

\begin{equation}
M_{ADM} = m_1+m_2 - \frac{1}{2 \pi} \int_\Omega\,\lim_{r \rightarrow \infty} \left(r^2 \frac{\partial u}{\partial r}\right) d\Omega, \label{eq12}
\end{equation} 

\noindent where $d \Omega = \sin \theta d\theta d\phi$.


\section{The Galerkin-Collocation decomposition method}%

We present here the GC domain decomposition algorithm to obtain initial data representing binary black holes. As the first step, we have divided the spatial domain into two subdomains denoted by $\mathcal{D}_1: 0 < r \leq r_0$ and $\mathcal{D}_2: r_0 \leq r < \infty$, where $r_0$ indicates the interface of these two non-overlapping subdomains. As a consequence, both subdomains share the same spherical angular coordinates $(\theta,\phi)$ which simplifies the implementation of the algorithm considerably.

The centerpiece of the algorithm is the spectral approximations of the regular functions $u^{(A)}(r,\theta,\phi)$ given by 

\begin{equation}
u^{(A)} = \sum^{N^{(A)}_x,N_y}_{k,l = 0}\sum^{l}_{m=-l}\,c^{(A)}_{klm} ~\chi^{(A)}_{k}(r) Y_{lm}(\theta,\phi). \label{eq13}
\end{equation}

\noindent where $A=1,2$ denotes the subdomains $\mathcal{D}_1,\mathcal{D}_2$, $c^{(A)}_{klm}$ represents the unknown coefficients or modes, $N^{(A)}_x$ and $N_y$ are, respectively, the radial and angular truncation orders that limit the number of terms in the above expansion. The angular patch has the spherical harmonics, $Y_{lm}(\theta,\phi)$, as the basis functions that are common to both domains. The choice of spherical coordinates together with the adoption of spherical harmonics basis functions are quite natural, and as we are going to show, are computationally very efficient and accurate. The radial basis functions $\chi^{(A)}_{k}(r)$ are defined following the prescription of the Galerkin method \cite{canuto,boyd}, in the sense of each function must satisfy the boundary conditions. Usually, they are obtained by taking suitable linear combinations of the Chebyshev polynomials as we are going to describe. 

Before defining the radial basis functions, it is necessary to introduce the computational subdomains. We have considered this feature an innovative part in constructing the algorithm. The Fig. 1 illustrates the mapping we have adopted. First, the entire radial domain $0 \leq r < \infty$ is mapped onto the interval $-1 \leq x < 1$ through the algebraic map \cite{boyd}

\begin{figure}
\includegraphics[scale=0.5]{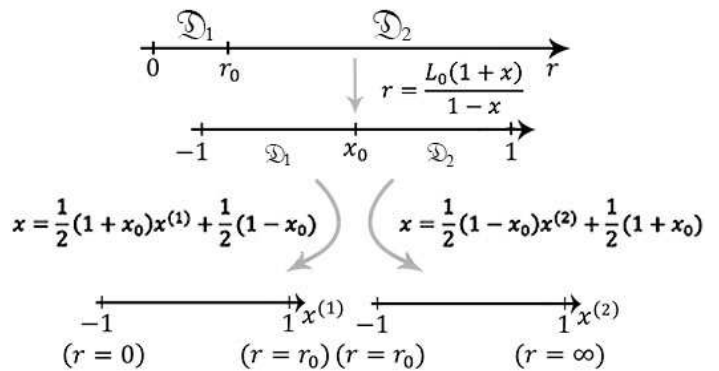}
\caption{Scheme showing the computational subdomains spanned by the coordinates $x^{(A)}$, $A=1,2$ and the corresponding maps that define them.}
\end{figure}

\begin{equation}
r = L_0\frac{(1+x)}{1-x}, \label{eq14}
\end{equation}

\noindent where $L_0$ is the map parameter. The subdomains $\mathcal{D}_1$ and $\mathcal{D}_2$ are now characterized by $-1 \leq x \leq x_0$ and $x_0 \leq x < 1$, respectively; $x_0$ is related to the interface radial coordinate $r_0$ by $r_0 = L_0(1+x_0)/(1-x_0)$. And second, we further define linear transformations $x^{(A)}=x^{(A)}(x)$, $A=1,2$ for the new computational domains for $\mathcal{D}_1$ and $\mathcal{D}_2$, respectively, such that $-1 \leq x^{(A)} \leq 1$ (cf. Fig. 1). The collocation points are designated by $x^{(A)}_k$ and mapped back to $r_k$ in the radial physical domain.

We are now in conditions to define the radial basis functions in each subdomain. With respect to $\mathcal{D}_1$, we define $\chi^{(1)}_k(r)$ by

\begin{equation}
\chi^{(1)}_k(r) = T_k\left(x^{(1)}=\frac{r+2L_0r/r_0-L_0}{r+L_0}\right), \label{eq15}
\end{equation}

\noindent where $T_k(x)$ is the Chebyshev polynomial of $k$th order, and $0 \leq r \leq r_0$ corresponds to $-1 \leq x^{(1)} \leq 1$. For the second domain $\mathcal{D}_2$, we have 

\begin{equation}
\chi^{(2)}_k(r) = \frac{1}{2} \left(TL_{k+1}(r)-TL_{k}(r)\right), \label{eq16}
\end{equation}

\noindent where $TL_k(r)$ is the redefined Chebyshev polynomial of $k$th order according to

\begin{eqnarray}
TL_k(r) = T_k\left(x^{(2)}=\frac{r-2r_0+{L_0}}{r+L_0}\right).  \label{eq17}
\end{eqnarray}

\noindent In this case the interval $r_0 \leq r < \infty$ is mapped out to $-1 \leq x^{(2)} < 1$. With the definition (\ref{eq15}) it can be shown that each basis function behaves asymptotically as $\chi^{(2)}_k(r) = \mathcal{O}(r^{-1})$. Therefore, we obtain the following asymptotic expression in the second domain

\begin{equation}
u^{(2)}(r,\theta,\phi) = \frac{\delta m(\theta,\phi)}{r} + \mathcal{O}\left(\frac{1}{r^2}\right), \label{eq18}
\end{equation}

\noindent where the function $\delta m(\theta,\phi)$ embodies the contribution to the ADM mass due to presence of angular and linear momenta. We can determine $\delta m(\theta,\phi)$ from

\begin{equation}
\delta m(\theta,\phi) = -\lim_{r \rightarrow \infty}\,r^2 \frac{\partial u^{(2)}}{\partial r}, \label{eq19}
\end{equation}

\noindent and the calculation of the ADM mass using Eq. (\ref{eq12}) becomes straightforward.

The spherical harmonics are complex functions implying that the coefficients $c^{(A)}_{klm}$ must be complex. Since the conformal factor is a real function, the real and imaginary parts of $c^{(A)}_{klm}$ satisfy the following symmetry relations

\begin{equation}
c^{(A)*}_{kl-m}=(-1)^{-m}\,c^{(A)}_{klm} \label{eq20}
\end{equation}

\noindent due to $Y^*_{l-m}(\theta,\phi)=(-1)^{-m}Y_{lm}(\theta,\phi)$. Consequently, the number of independent coefficients in each domain is $\left(N^{(A)}_x+1\right)\left(N_y+1\right)^2$.

We have to guarantee that the spectral approximations of $u^{(1)}(r,\theta,\phi)$ and $u^{(2)}(r,\theta,\phi)$ given by expression (\ref{eq13}) represent the same function at the match point of the domain. This is done by imposing the continuity at the interface $r=r_0$ that separates both subdomains through the following matching conditions

\begin{eqnarray}
u^{(1)}(r_0,\theta,\phi)&=&u^{(2)}(r_0,\theta,\phi) \nonumber \\
\\
\left(\frac{\partial u^{(1)}}{\partial r}\right)_{r=r_0}&=&\left(\frac{\partial u^{(2)}}{\partial r}\right)_{r=r_0}.  \nonumber \label{eq21}
\end{eqnarray}

We now establish the residual equation associated with the Hamiltonian constraint in each domain by substituting the spectral approximations represented by Eq. (\ref{eq13}) into the Hamiltonian constraint (\ref{eq7}). In addition, we have taken into account the differential equation for the spherical harmonics to get rid of the derivatives with respect to $\theta$ and $\phi$. We have arrived to the following expression

\begin{widetext}
\begin{eqnarray}
\mathrm{Res}^{(A)}(r,\theta,\phi)=\sum_{k,n,p}\,c^{(A)}_{knp}\Big(\frac{1}{r^2}\frac{\partial}{\partial r}\left(r^2 \frac{\partial \chi^{(A)}_k}{\partial r}\right)- \frac{n(n+1)}{r^2}\chi^{(A)}_k \Big) Y_{np}(\theta,\phi) + \frac{\left(\bar{A}^{ij}\bar{A}_{ij}\right)^{(A)}}{8\left[u^{(A)}(r,\theta,\phi)+\frac{1}{2}\left(\frac{m_1}{r_{C_1}}+\frac{m_2}{r_{C_2}}\right)\right]^{7}} \label{eq22}
\end{eqnarray}
\end{widetext}

\noindent with $A=1,2$ corresponding to the domains $\mathcal{D}_1$ and $\mathcal{D}_2$, respectively. 

The next and final step is to describe the procedure to obtain de coefficients $c^{(A)}_{klm}$. We have followed the implementation of one domain \cite{clemente} straightforwardly. From the method of weighted residuals \cite{finlayson}, these coefficients are evaluated with the condition of forcing the residual equation to be zero in an average sense. It means that 

{\small{
\begin{eqnarray}
&& \left<\mathrm{Res}^{(A)},R^{(A)}_j(r)S_{lm}(\theta,\phi)\right> \equiv \nonumber \\
\nonumber \\
&& \int_{\mathcal{D}^{(A)}}\,\mathrm{Res}^{(A)} R^{(A)*}_j(r) S^{*}(\theta,\phi)_{lm}\,w_r^{(A)} w_\theta w_\phi \,r^2 dr d\Omega=0,\nonumber\\   \label{eq23}
\end{eqnarray}
}}

\noindent where the functions $R^{(A)}_j(r)$ and $S_{lm}(\theta,\phi)$ are called the test functions while $w_r^{(A)}, w_\theta$ and $w_\phi$ are the corresponding weights. In both domains we choose the radial test function as prescribed by the Collocation method \cite{boyd,fornberg}:

\begin{equation}
R^{(A)}_j(r) = \delta\left(r-r^{(A)}_j\right), \label{eq24}
\end{equation}

\noindent which is the delta of Dirac function, $r^{(A)}_j$, $A=1,2$, represents the collocation points defined in each domain and $w_r^{(A)}=1$. Following the Galerkin method we identify the angular test function $S_{lm}(\theta,\phi)$ as the spherical harmonics and as a consequence $w_\theta=w_\phi=1$. Therefore Eq. (\ref{eq23}) becomes

\begin{equation}
\left<\mathrm{Res}^{(A)}(r,\theta,\phi),Y_{lm}(\theta,\phi)\right>_{r=r^{(A)}_j}=0, \label{eq25}
\end{equation}

\noindent where $A=1,2$, $l=0,1,..,N_y$ and $m=0,1,..,l$. As we are going to show, the number of radial collocation points defined in each domain provides the correct number of equations for the modes.

Before going further, we need to consider the matching conditions in respect to the approximation adopted above. The corresponding residuals $\mathrm{Res}_1(\theta,\phi)=u^{(1)}(r_0,\theta,\phi)-u^{(2)}(r_0,\theta,\phi)$ and $\mathrm{Res}_2(\theta,\phi)=(\partial u^{(1)}/\partial r - \partial u^{(2)}/\partial r)_{r=r_0}$ are approximated as

{\small
\begin{eqnarray}
& & \left<\mathrm{Res}_1,Y_{lm}\right>=\sum_{k=0}^{N^{(1)}_x} c^{(1)}_{klm}\chi^{(1)}_k(r_0)-\sum_{k=0}^{N^{(2)}_x} c^{(2)}_{klm}\chi^{(2)}_k(r_0) =0  \nonumber \\
\nonumber \\
& & \left<\mathrm{Res}_2,Y_{lm}\right>=\sum_{k=0}^{N^{(1)}_x} c^{(1)}_{klm}\left(\frac{d \chi^{(1)}_k}{d r}\right)_{r_0}-\sum_{k=0}^{N^{(2)}_x} c^{(2)}_{klm}\left(\frac{d \chi^{(1)}_k}{d r}\right)_{r_0} = 0, \nonumber \\ \label{eq26}
\end{eqnarray}
}

\noindent where $l=0,1,..,N_y$ and $m=0,..,l$. Notice that these expressions result from exact integrations on the angular domain (cf. Eq. (\ref{eq23})) due to the orthogonality of the spherical harmonics. Thus, we ended up with $2 (N_y+1)^2$ linear relations of the coefficients of both domains. 

\begin{figure}[htb]
\includegraphics[scale=0.5]{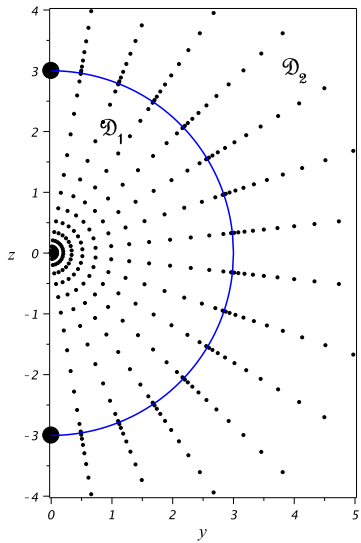}
\caption{Illustration of the collocation points in both domains projected into the plane $yz$. The black circles located along the axis $z$ represents the punctures and blue semicircle, $r_0=a$, is the interface of both domains.}
\end{figure}

At this point we present the radial collocation points at each domain. By taking into account the matching conditions, we have $(N_x^{(1)}+N_x^{(2)})(N_y+1)^2$ unknown coefficients in both domains, therefore it is necessary the same number of equations resulting from Eq. (\ref{eq25}). Thus, we need $N_x^{(A)}$ radial collocation points in each computational subdomain given by


\begin{eqnarray}
x_j=\cos\left[\frac{(2j+1) \pi}{2 N_x^{(A)}}\right],\;\;j=0,1,..,N_x^{(A)}-1, \label{eq27}
\end{eqnarray}

\noindent and the corresponding radial points in the corresponding physical subdomain are

\begin{eqnarray}
r^{(1)}_j &=& \frac{L_0 (1+x_j)}{2L_0/r_0 + 1-x_j},\, \mathrm{domain}\,\,\mathcal{D}_1 \label{eq28}\\
\nonumber \\
r^{(2)}_j &=& \frac{2r_0+L_0(1+x_j)}{1-x_j},\, \mathrm{domain}\,\,\mathcal{D}_2 \label{eq29}
\end{eqnarray}
 
\noindent with $j=0,2..,N_x^{(A)}-1$. We remark that the point at infinity is excluded since the residual equation is automaticaly satisfied asymptotically due to the choice of the radial basis function (\ref{eq16}). Noticed that the origin is also excluded. For the sake of illustration, we show in Fig. 2 the organization of the radial collocation points in both subdomains.

We are in conditions to present a more detailed form of the set of equations represented by Eq. (\ref{eq22}), after using the orthogonality of the spherical harmonics in the first two terms of the residual equation (\ref{eq25}):

{\small
\begin{widetext}
\begin{eqnarray}
\left<\mathrm{Res}^{(A)},Y_{lm}(\theta,\phi)\right>_{r_j} = \sum_{k}\,\frac{c^{(A)}_{klm}}{r_j^2}\left(\frac{\partial}{\partial r}\left(r^2 \frac{\partial \chi^{(A)}_k}{\partial r}\right) - l(l+1)\chi^{(A)}_k \right)_{r_j} + \left<\frac{\left(\bar{A}^{ij}\bar{A}_{ij}\right)^{(A)}}{8\left[u^{(A)}(r,\theta,\phi)+\frac{1}{2}\left(\frac{m_1}{r_{C_1}}+\frac{m_2}{r_{C_2}}\right)\right]^{7}},Y_{lm}(\theta,\phi)\right>_{r_j}=0, \nonumber \\ \label{eq30}
\end{eqnarray}
\end{widetext}
}

\noindent where $l=0,1,..,N_y$, $m=0,1,..,l$ and $j=0,1,..,N_x^{(A)}$. Therefore, we have obtained a total $(N^{(1)}_x+N^{(2)}_x)(N_y+1)^2$ nonlinear algebraic equations, that together with the $2(N_y+1)^2$ equations from the matching conditions (\ref{eq26}), constitute the set of equations to be solved for the modes $c_{klm}^{(A)}$. As a final piece of information, we have calculated the last term of the above equation using quadrature formulae as indicated below 


\begin{eqnarray}
\left<(..),Y_{lm}(\theta,\phi)\right>_{r_j} = \sum_{k,n=0}^{N_1,N_2}\,(..) Y^*_{lm}(\theta_k,\phi_n) v^{\theta}_k v^{\phi}_n, \label{eq31} 
\end{eqnarray}

\noindent where $(\theta_k,\phi_n)$, $k=0,1,..,N_1$, $n=0,1,..,N_2$ are the quadrature collocation points, and $v^{\theta}_k v^{\phi}_n$ are the corresponding weights \cite{fornberg}. For better accuracy we have set $N_1=N_2=2 N_y+1$, but this is not mandatory since it is possible to use simply $N_1=N_2=N_y$. 

In closing this Section, it is useful to comment on the possibility of damaging the exponential convergence of the numerical solution to the Hamiltonian constraint due to the singularities of the punctures. The first is to choose another form of the conformal factor with the requirement of being regular everywhere as established by the moving puncture method and applied to the initial data problem in connection with trumped black holes \cite{baumgarte_12,dietrich}. Alternatively, it is possible to set suitable coordinates in which the puncture are located at the edge of the computational domain, or as adopted in Refs. \cite{ansorg_05,ansorg_07} in placing the punctures at the domain interface. We have followed the latter approach (cf. Fig. 2) avoiding to coincide any collocation point coinciding with the loci of the punctures. As we are going to show in the next Section, the exponential convergence in all examples.

\section{Numerical implementation}%

The computational framework was implemented initially in Maple. The Maple quasi-numerical script was used as a reference to develop a serial code in Fortran. Although algorithmically different,  both programs produced the same output for monitored variables. This constituted an excellent validation of the Fortran code. In this work, we had used only the numerical code in Fortran when the memory and the velocity were a real limit for simulations. In turn, the Fortran solver allowed us to identify the stage that needed more computational resources. Thus, we have implemented a parallelization to proceed with the determination of the desired solution. 

Our numerical problem has $(N_x^{(1)} + N_x^{(2)})(N_y+1)^2$ nonlinear equations from the Hamiltonian constraint (\ref{eq30}) and $2(N_y+1)^2$ linear equations from the matching conditions (\ref{eq26}). These three sets of equations are deployed in one vector $H_n(u)=0$, with $n=1...N_z$, and $N_z=(N_x^{(1)} + N_x^{(2)}+2)(N_y+1)^2$. The system of equations $H_n$ has $z_n$ solutions; each solution $z_n$ corresponds to one and only one real (or imaginary) part of the coefficient $c^{(A)}_{jkl}$.

To solve the system of equations, we use the standard Newton-Raphson method \cite{NumericalRecipes}

\begin{equation}
J\delta z_n = - H_n, \label{eq32}
\end{equation}

\noindent where $J=(\partial H_n/\partial z_m)$ is the Jacobian matrix and $\delta z$ is the variation of the solution $z_n$ between the iteration $N_i$ and $N_{i-1}$, up to some specified tolerance for the convergence. We stress here the fact that the Jacobian is calculated numerically using forward finite differences with excellent results. To get the set of solutions at each iteration, we use an LU or QR decomposition \cite{NumericalRecipes}. We observe the best performance for the LU decomposition. All the special functions and its derivatives are calculated using the standard generating formulae from the Numerical Recipes library of subroutines.

\section{Numerical Results}%


We have considered three examples of initial data of binary black holes to show the fast convergence of the domain decomposition Galerkin-Collocation algorithm. We have started with an axisymmetric configuration of boosted black holes and in the sequence, two distinct three-dimensional binary systems formed with black holes with angular and linear momenta. 

\begin{figure}[htb]
{\includegraphics[scale=0.23]{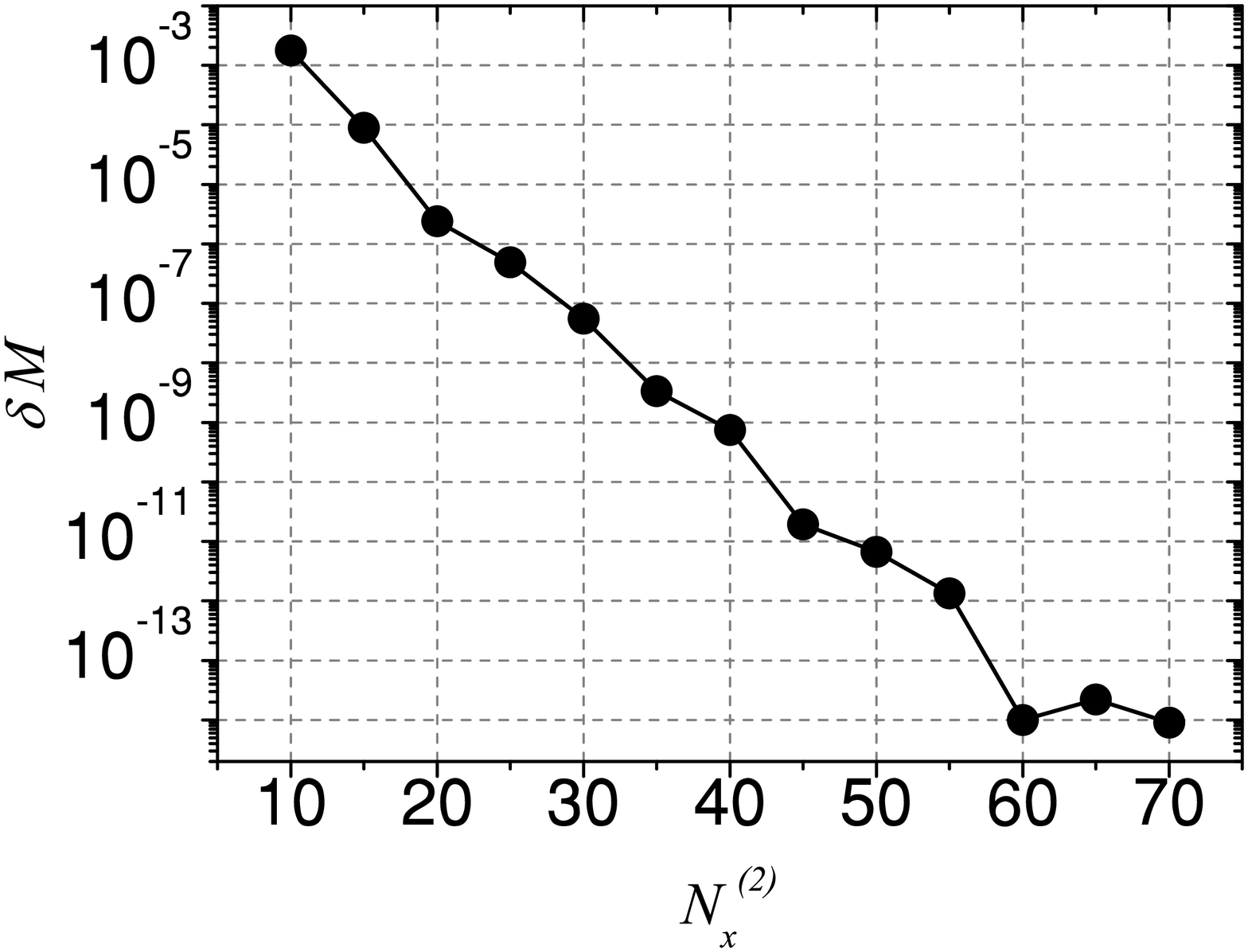}}
{\includegraphics[scale=0.23]{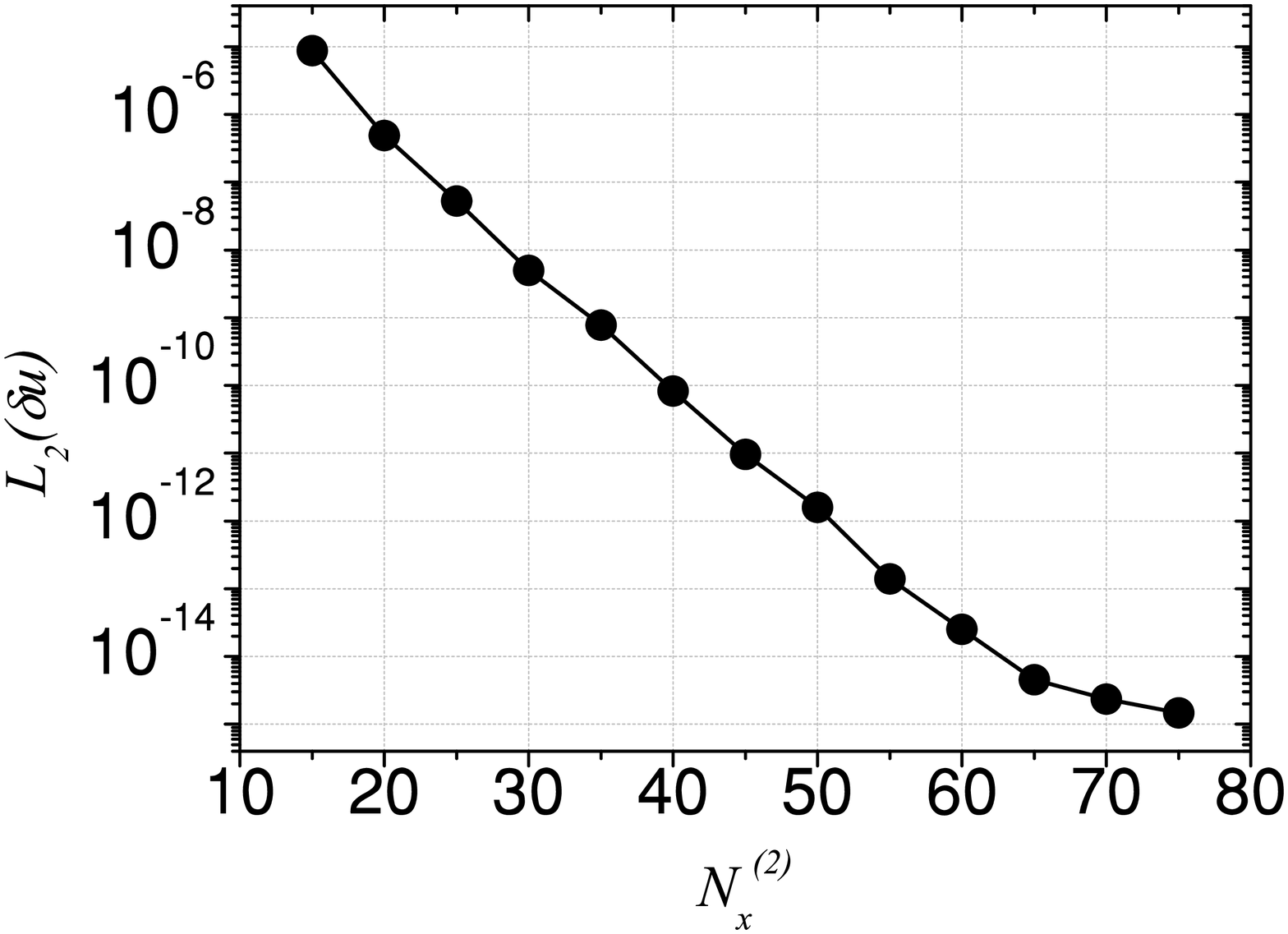}}
\caption{Convergence of the ADM mass (top panel) and of the $L_2$-error $L(\delta u)$ (bottom panel) for the example of axisymmetric binary black holes. In both cases, the exponential convergence is achieved.}
\end{figure}


The first example consists of two boosted black holes represented by  punctures of equal masses $m_1=m_2=0.5 m$ and $m=m_1+m_2=1.0$.  The punctures are placed on the axis $z$ at $\mathbf{r}_1=(0,0,-a)$ and $\mathbf{r}_1=(0,0,a)$, respectively, where $2 a$ is the coordinate separation between the punctures. The corresponding linear  momenta are $\mathbf{P}_1=(0,0,P_1)$ and $\mathbf{P}_2=(0,0,P_2)$ which yields the following expression for source-term of the Hamiltonian: 

{\small 
\begin{eqnarray}
&&\bar{A}_{ij}\bar{A}^{ij} = \frac{9 P_1^2}{2 r_1^6} \left[(1+2\cos^2\theta)r^2 + 6ar\cos\theta + 3a^2\right] \nonumber \\
&&+ \frac{9 P_2^2}{2 r_2^6} \left[(1+2\cos^2\theta)r^2 - 6ar\cos\theta + 3a^2\right] + \frac{9 P_1 P_2}{2 r_1^5 r_2^5} \times \nonumber \\
 &&\big[(1+2\cos^2\theta)r^6+ (2\cos^4\theta-14\cos^2\theta+3)a^2r^4 + \nonumber \\
&&(8\cos^2\theta+1)a^4r^2 - 3a^6\big],\label{eq33}
\end{eqnarray}}
 


\noindent where $r_{1,2}=\sqrt{r^2 \pm 2 a r \cos \theta +a^2}$. Notice that since $\bar{A}_{ij}\bar{A}^{ij} $ does not depend on the angle $\phi$; we have an axisymmetric configuration. In this case, the Legendre polynomials replace the spherical harmonics as the angular basis functions in the spectral approximation of Eq. (\ref{eq13}).  Alternatively, we can translate the axisymmetry in the spectral representation by $c^{(A)}_{jkl}=0$ for all $l \neq 0$.  

For the numerical convergence tests, we have chosen  $P_1=-P_2=P_0=0.2m$, $a=3.0m$ and fix $N_y=16$, $N_x^{(1)}=15$. Then, we have proceeded by varying the radial truncation order of the second domain as $N_x^{(2)}=5,10,15,...$ and solved the system of Eqs. (\ref{eq26}) and (\ref{eq30}) for each $N_x^{(2)}$ . With the modes $c^{(A)}_{jkl}$ determined, the ADM mass is calculated according to Eqs. (\ref{eq12}) and (\ref{eq19}). In the sequence, we have evaluated the difference between the ADM masses corresponding to successive solutions through $\delta M(N_x^{(2)}) = |M_{ADM}(N_x^{(2)}+5)-M_{ADM}(N_x^{(2)})|$. Fig. 3 shows the exponential decay of $\delta M$ indicating the rapid convergence of the numerical solution. Note that for $N_x^{(2)} \geq 60$ the saturation due to round-off error is achieved in about $10^{-14}$. In these numerical experiments, we have chosen $r_0=L_0=a$ for the interface and the map parameter. The ADM mass of the axisymmetric binary of boosted black hole is $M_{ADM} \approx 1.06612795$. 

Another convergence test is provided by the decay of the $L_2$-error between two successive solutions of the regular function $u^{(2)}(r,\theta)$ obtained previously. We have calculated the $L_2$-error using the following expression

\begin{eqnarray}
L_2(\delta u) = \sqrt{\frac{1}{8 \pi} \int_0^{2\pi}\,\int_{-1}^{1}\,\int_{-1}^{1}\,(\delta u)^2 d\phi d\bar{y} dx^{(2)}}, \label{eq34}
\end{eqnarray}

\noindent where $\delta u = u^{(2)}_{N_x^{(2)}+5}-u^{(2)}_{N_x^{(2)}}$ and  $\bar{y} = \cos \theta$. From Fig. 3, the exponential convergence is achieved similarly to the convergence of the ADM M mass.

The second example is the three-dimensional boosted binary of black holes studied by Ansorg et al. \cite{ansorg_04}. The punctures have the same masses $m_1=m_2=0.5m$ with $m=m_1+m_2=1.0$, and lie on the axis $x$ at $\mathbf{r}_1=(-a,0,0)$ and $\mathbf{r}_2=(a,0,0)$. The linear momenta of the punctures are $P_1=(0,P_1,0)$ and $P_2=(0,P_2,0)$ and the corresponding source-term $\bar{A}_{ij} \bar{A}^{ij}$ becomes 

{\small
\begin{eqnarray}
\bar{A}_{ij}\bar{A}^{ij} &=& \frac{9 P_2^2}{2 r_2^6}(a^2+r^2+2 a r \sin \theta \cos \phi + 2 r^2 \sin^2 \theta \sin^2 \phi) \nonumber \\
&+&\frac{9 P_2^2}{2 r_2^6}(a^2+r^2-2 a r \sin \theta \cos \phi + 2 r^2 \sin^2 \theta \sin^2 \phi) \nonumber \\
&+&\frac{9 P_1 P_2}{2 r_1^3 r_2^3}\Big[r^2-a^2+\frac{2r^2 \sin^2\theta\,\sin^2 \phi}{r_1^2 r_2^2}(r^4-a^2r^2-a^4 \nonumber \\
&+&a^2r^2\sin^2\theta\,\cos^2\phi)\Big], \label{eq35}
\end{eqnarray}
}

\noindent where $r_{1,2}=\sqrt{r^2 \pm 2ar\,\sin \theta\,\cos \phi + a^2}$. With $P_2=-P_1=0.2m$,  we ended up with a particular expression that induces additional symmetries for the coefficients $c^{(A)}_{jkl}$, meaning that some of the $(N_x^{(1)}+N_x^{(2)})(N_y+1)^2$ coefficients vanishes. In this case, we have found that all imaginary parts of the coefficients vanish, $\mathrm{Im}(c^{(A)}_{jkl})=0$, and some of the real coefficients according to $\mathrm{Re}(c^{(A)}_{jkl})=0$ if $k-l$ is an odd number.  We have performed the convergence tests of the ADM mass and the $L_2(\delta u)$ error associated with the difference between two successive solutions in the second domain. The results depicted in Fig. 4 present similar exponential decay suggesting we can select one of the tests to verify the convergence of the code. In the numerical experiments, we have set as before $r_0=L_0=a_0$, and the ADM mass of the three-dimensional boosted binary system is $M_{ADM} \simeq 1.065895065$. Ansorg et al. \cite{ansorg_04} have used a grid of $98 \times 98 \times 50 = 480,200$ points, whereas we have used a total of 920 coefficients in the highest resolution of $N_y=6, N_x^{(1)}=15, N_x^{(2)}=75$. Note that with these values the total number of coefficients would be $(N_x^{(1)}+N_x^{(2)}+2)(N_y+1)^2=4,508$, but the additional symmetries have reduced it drastically, moreover we have required a total of $(N_x^{(1)}+N_x^{(2)}+2)(2N_y+2)^2=18,032$ grid points (here $N_1=N_2=2N_y+1$ for the quadrature formulae (\ref{eq31})).


\begin{figure}[htb]
{\includegraphics[scale=0.23]{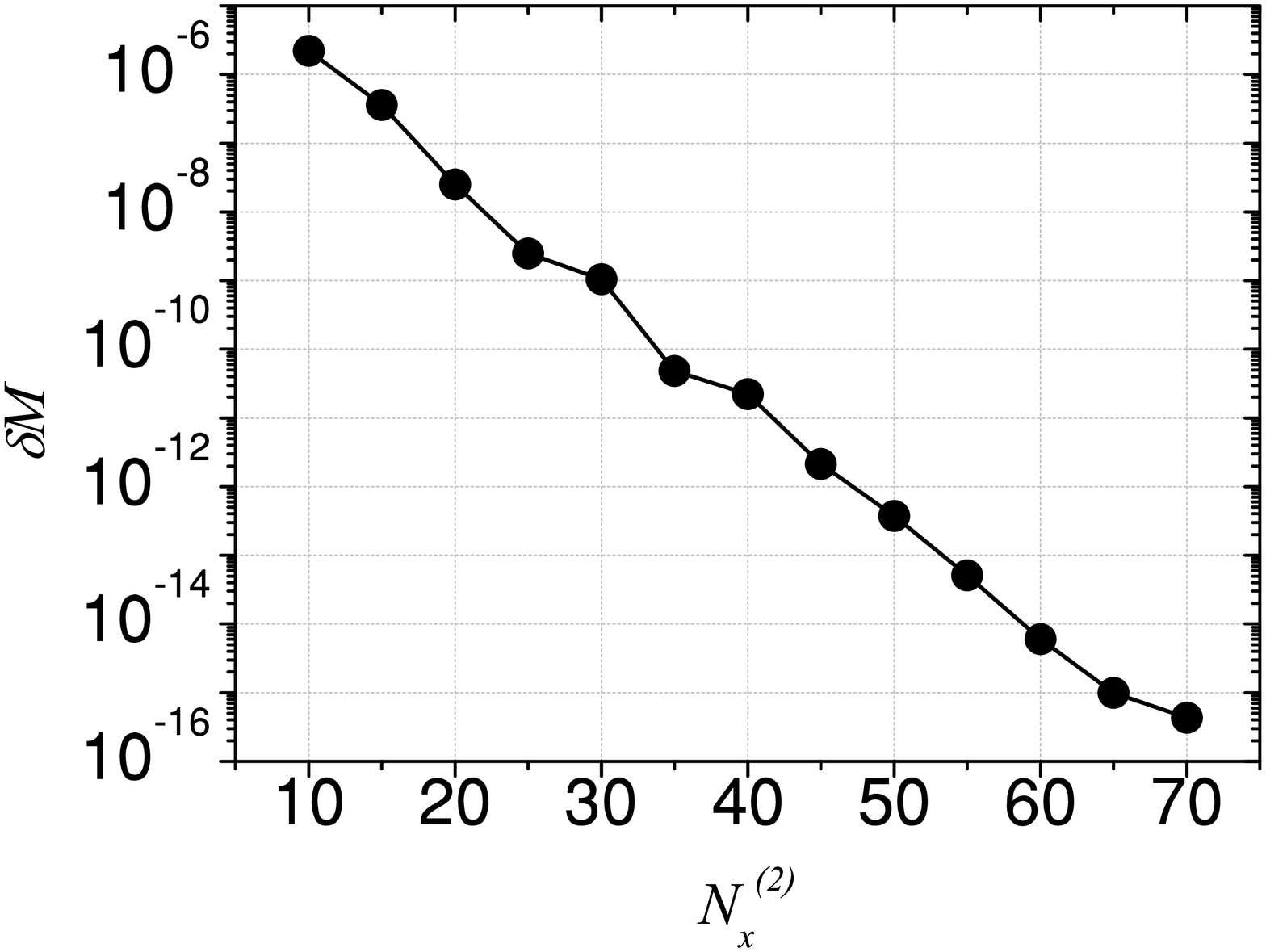}}
{\includegraphics[scale=0.23]{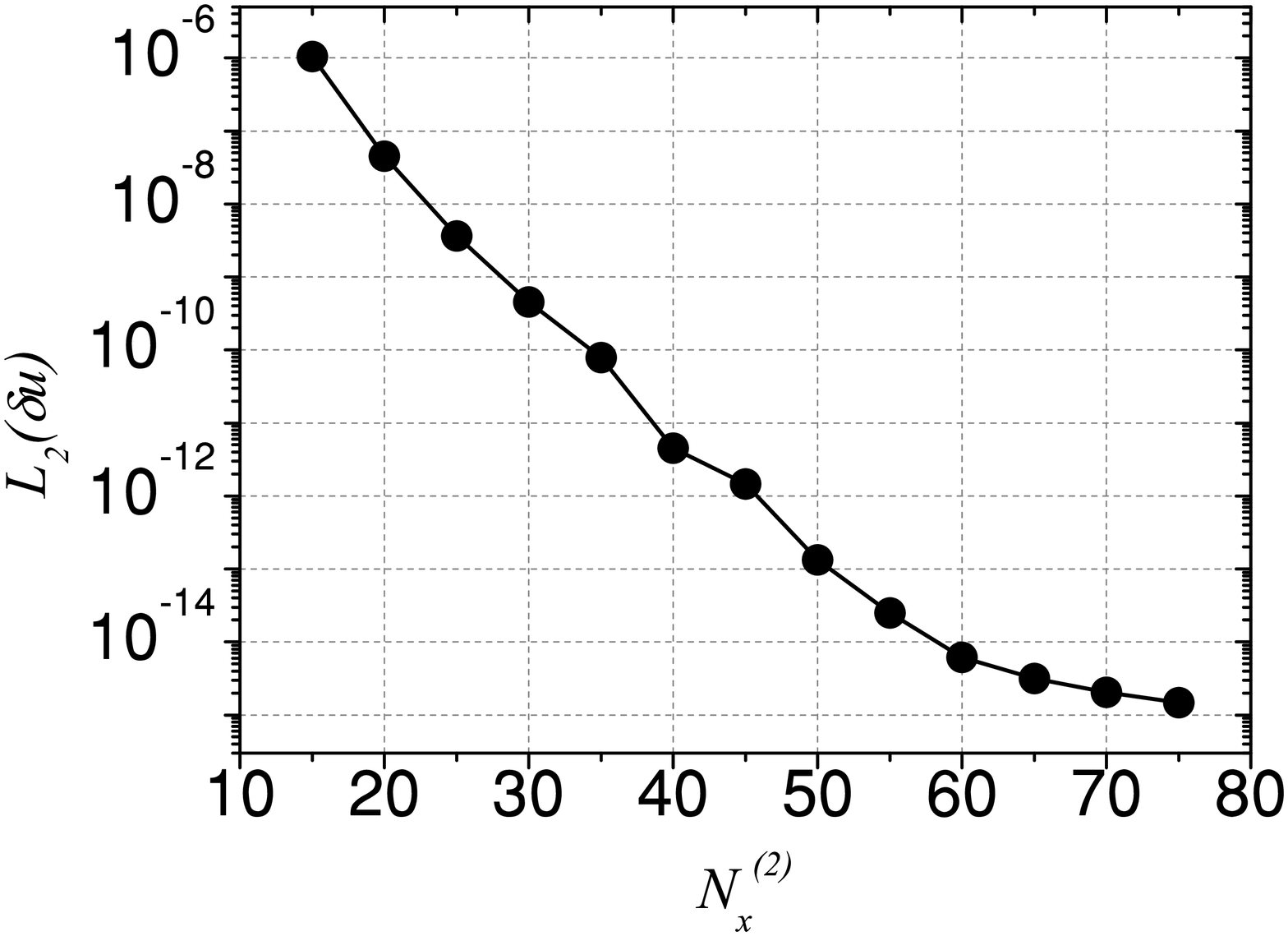}}
\caption{Convergence of the ADM mass (top panel) and of the $L_2$-error $L(\delta u)$ (bottom panel) for the second example of a non-axisymmetric binary black hole. In both cases, the exponential convergence is achieved.}
\end{figure}

The last example is a general three-dimensional initial data of a binary of spinning-boosted black holes. This initial configuration is taken from Br\"ugmann \cite{brugman_99} in which the puncture masses are $m_1=1.5,\,m_2=1.0$, located at $\mathbf{r}_1=(0,0,-1.5)$ and $\mathbf{r}_2=(0,0,1.5)$, respectively. The linear and angular momenta of each puncture are $\mathbf{P}_1=(P_0,0,0)$, $\mathbf{P}_2=(-P_0,0,0)$, $\mathbf{S}_1=(-S_0,S_0,0)$ and $\mathbf{S}_2=(0,2 S_0,2 S_0)$, where $P_0=2.0$ and $S_0=0.5$.

We have omitted writing here the long expression of the source-term $\bar{A}^{ij}\bar{A}_{ij}$, but contrary to the previous examples there are no additional symmetries reduce the number of independent coefficients $c^{(A)}_{jkl}$ other than expressed by Eq. (\ref{eq20}). Further, we have restricted to verify the convergence of the ADM mass evaluating the quantity $\delta M$ as already established, remarking that we have set $N_y=6$ and $N_x^{(1)}=10$. Fig. 5 shows a clear exponential decay of $\delta M$ until $N_x^{(2)}=45$. To complete all the pertinent information for the numerical experiments, we have set $r_0=L_0=a$ as in the other two examples, and the calculated ADM mass is $M_{ADM} \simeq 3.07765268$. It seems that these choices for the location of the interface as well the map parameter in the present code is general for any binary black hole system. This system was solved using the BAM code \cite{bam_brug} with a grid of $65^3=21,125$ points, while in the present algorithm we have required the maximum of $(N_x^{(1)}+N_x^{(2)}+2)(N_y+1)^2=2,793$ coefficients in both domains for $N_y=6, N_x^{(1)}=10, N_x^{(2)}=45$. For the sake of illustration, after expressing the regular function in terms of the cartesian coordinates, i.e. $u^{(A)}=u^{(A)}(\hat{x},\hat{y},\hat{z})$, we have projected $u^{(A)}-1$ into the plane $\hat{x}=0$ as shown in Fig. 6. Notice the asymmetry due to the distinct spins of both black holes.

\begin{figure}[htb]
\includegraphics[height=5.cm,width=6.0cm]{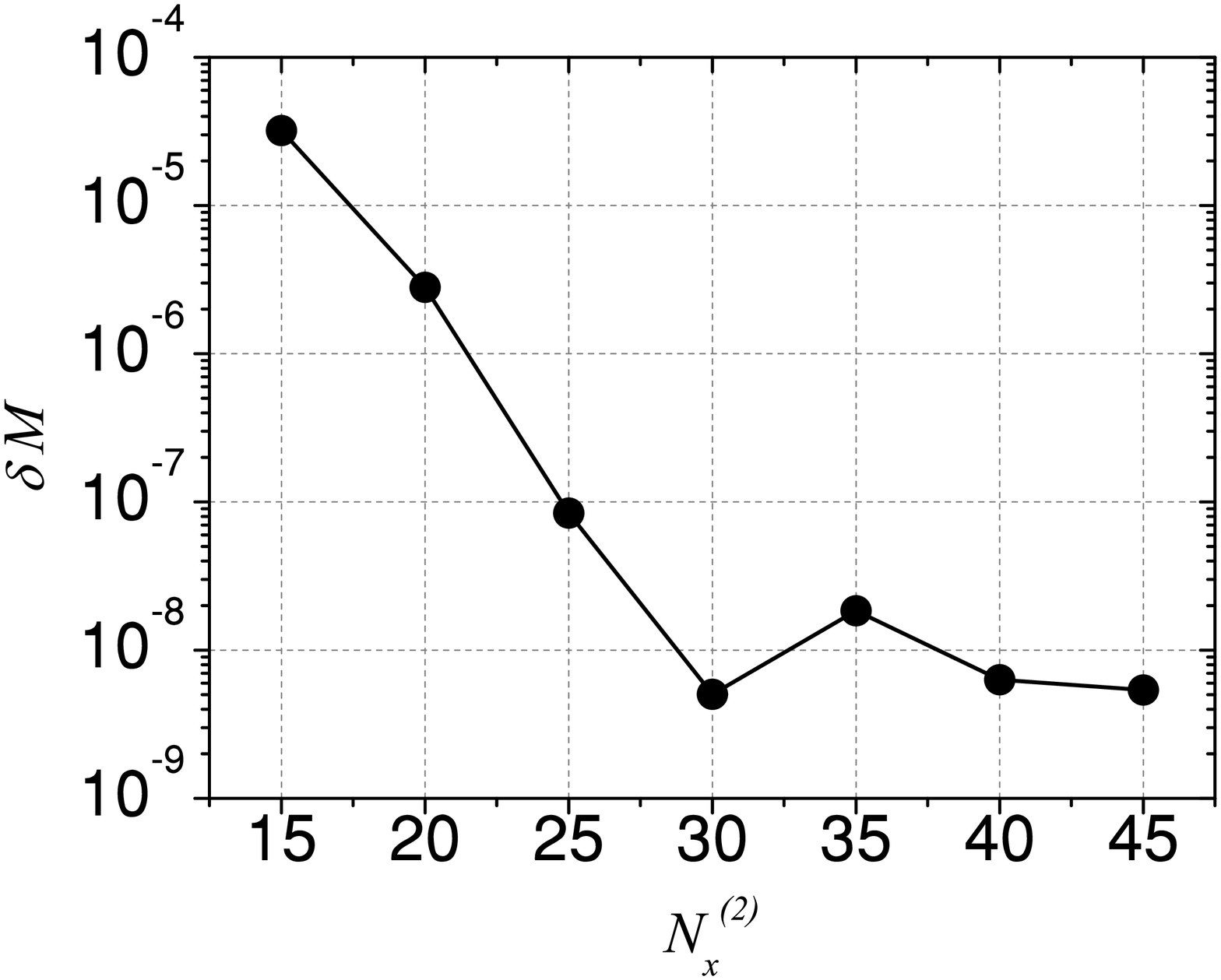}
\caption{Convergence of the ADM mass for the binary of boosted-spinning black holes of the third example.}
\end{figure}

\begin{figure}[htb]
\includegraphics[scale=0.55]{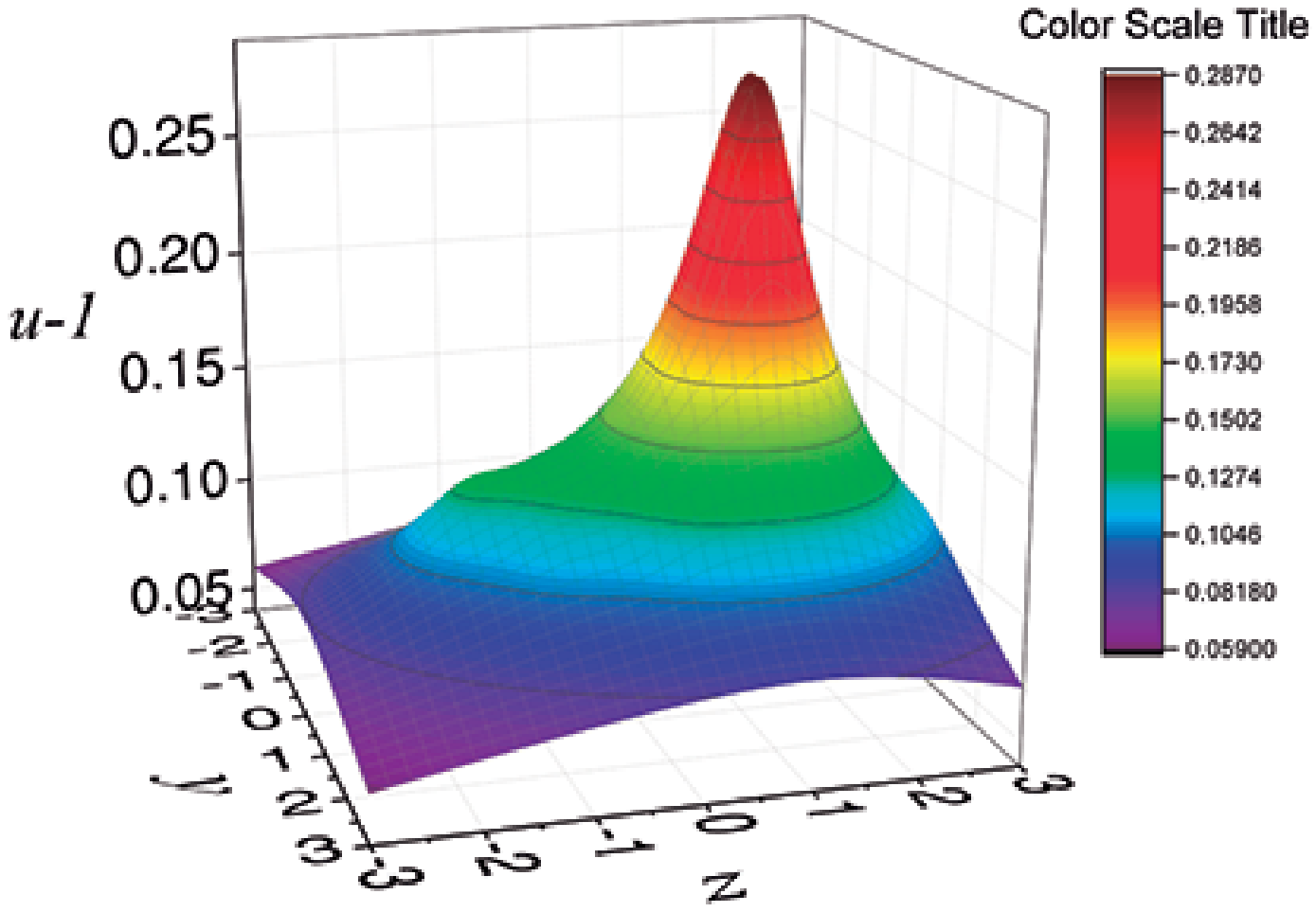}
\includegraphics[height=6.cm,width=8.0cm]{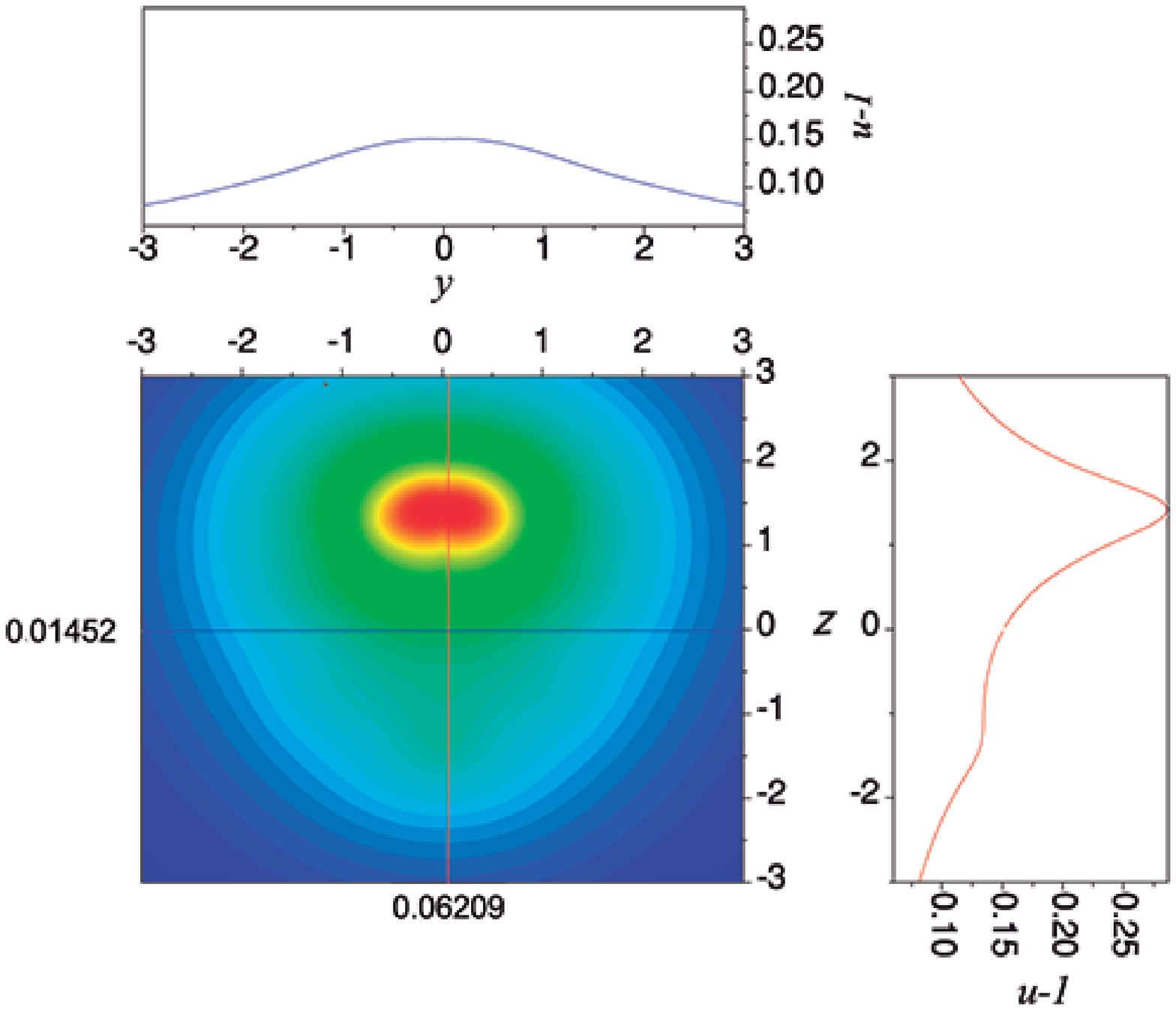}
\caption{(Top panel) Three-dimensional plot of $u(x=0,y,z)-1$ where $(x,y,z)$ are the cartesian coordinates. The punctures are located along the $z$-axis at $z=\pm a=\pm 0.6m$, with $m=m_1+m_2=2.5$. The assymetry is due to the distinct spin component of each black hole.(Bottom panel) Projections of $u(y,z)-1$ into the planes $z=\mathrm{constant}$, $u=\mathrm{constant}$ and $y=\mathrm{constant}$.}
\end{figure}

\section{Concluding remarks and outlook of future work}

We have presented a new DD code base on the GC method for the Bowen-York initial data representing arbitrary binary black hole systems. It is worth of mentioning some of the leading aspects of the algorithm that makes it simpler and distinct from other numerical procedures. 

In the present algorithm, we have covered the whole spatial domain with the spherical coordinates no matter which binary system is under consideration. We have split the spatial domain into two subdomains defined by $\mathcal{D}_1: 0 \leq r \leq r_0$ and  $\mathcal{D}_2: r_0 \leq r < \infty$, such that the angular coordinates of the collocation points are the same for both subdomains as indicated by Fig. 2. As the central piece of the code, we have provided the spectral approximation of the regular component of the conformal factor in each subdomain, $u^{(A)}(r,\theta,\phi), A=1,2$, with the spherical harmonics as the angular basis functions. The radial basis functions are defined in each subdomain to satisfy the boundary conditions. 

We have solved the Hamiltonian constraint (\ref{eq7}) for three distinct BBH represented by punctures located on the $z$ axes. The first is an axisymmetric boosted binary; the second example is a three-dimensional binary boosted taken from Ansorg et al. \cite{ansorg_04}. In both cases the puncture masses are equal.  The third example is an arbitrary spinning-boosted binary of distinct puncture masses drawn from Ref. \cite{brugman_99}. The tests we have employed to validate the code was the convergence of the ADM mass and the regular function $u^{(2)}(r,\theta,\phi)$. As expected the convergence is exponential as shown by the Figures. 

The details of the code performance and its parallelization for massive computations will be addressed elsewhere. Meanwhile, we only mention here that the first two binary systems were solved using one thread for computations. In the last example, however, we required multi-threading to deal with the Hamiltonian constraint.  

The present algorithm can be applied to some problems. The most relevant is the obtention of high-resolution initial data for a binary system of neutron stars. For the BBH system, we can straightforwardly consider trumpets instead wormholes representations. Moreover, we could use the conformal thin sandwich approach for the initial data. These are just some possible directions and venues for our coming work, including evolutions.  

\begin{acknowledgements} 
The authors acknowledge the financial support of the Brazilian agencies Conselho Nacional de Desenvolvimento Cient\'{i}fico e Tecnol\'{o}gico (CNPq) and Funda\cao\ Carlos Chagas Filho de Amparo \`{a} Pesquisa do Estado do Rio de Janeiro (FAPERJ). H. P. O. thanks FAPERJ for support within the Grant No. E-26/202.998/518 2016 Bolsas de Bancada de Projetos (BBP).
\end{acknowledgements}

\end{document}